\documentclass[aip,rsi,graphicx,reprint]{revtex4-1}

\usepackage{graphicx}
\usepackage{dcolumn}
\usepackage{bm}
\usepackage{float}

\usepackage[utf8]{inputenc}
\usepackage[T1]{fontenc}
\usepackage{mathptmx}
\usepackage{etoolbox}
\usepackage{physics}

\usepackage{epsfig} 

\usepackage{epstopdf}
\usepackage{multirow}
\usepackage{amsmath}
\usepackage{xcolor}
\usepackage{gensymb}
\usepackage{kantlipsum}  
\usepackage{hyperref}
\usepackage{braket}
\usepackage{soul,xcolor}

\begin{document}
\setstcolor{black}

\title{A planar cloverleaf antenna for the creation of circularly polarized microwave fields} 
\author{Weijun Yuan}
\thanks{These authors contributed equally.}
\affiliation{Department of Physics, Columbia University, New York, New York 10027, USA}
\author{Siwei Zhang}
\thanks{These authors contributed equally.}
\affiliation{Department of Physics, Columbia University, New York, New York 10027, USA}
\author{Niccol\`{o} Bigagli}
\affiliation{Department of Physics, Columbia University, New York, New York 10027, USA}
\author{Claire Warner}
\affiliation{Department of Physics, Columbia University, New York, New York 10027, USA}
\author{Ian Stevenson}
\affiliation{Department of Physics, Columbia University, New York, New York 10027, USA}
\author{Sebastian Will}
\email{ sebastian.will@columbia.edu}
\affiliation{Department of Physics, Columbia University, New York, New York 10027, USA}
\date{\today}

\begin{abstract}
We report on the design and characterization of a compact microwave antenna for atomic and molecular physics experiments. The antenna is comprised of four loop antennas arranged in cloverleaf shape, allowing for precise adjustment of polarization by tuning the relative phase of the loops. We optimize the antenna for left-circularly polarized microwaves at 3.5 GHz and characterize its performance using ultracold NaCs molecules as a precise quantum sensor. Observing an unusually high Rabi frequency of $2\pi \times 46$ MHz, we extract an electric field amplitude of 33(2)~V/cm at 22 mm distance from the antenna. The polarization ellipticity is 2.3(4)$\degree$, corresponding to a 24~dB suppression of right-circular polarization. The cloverleaf antenna is planar and provides large optical access,  making it highly suitable for quantum control of atoms and molecules, and potentially other quantum systems that operate in the microwave regime.
\end{abstract}


\maketitle 

\section{Introduction}

Microwave fields play a key role in modern technology. In everyday life, numerous applications rely on the emission and detection of microwaves, from microwave ovens to wireless data communication.~\cite{ishii1995handbook,ohlsson2001microwave,golio2018rf} In quantum science, the active use of microwaves dates back to the 1930s, when for the first time a nuclear quantum spin was flipped by applying an oscillating microwave field.~\cite{rabi1937space} Recently, the importance of microwave fields in quantum science has rapidly risen. Many high-quality qubit systems operate in the microwave regime, including superconducting qubits,~\cite{blais2004cavity} nitrogen-vacancy (NV) centers,~\cite{awschalom2018quantum} quantum  dots,~\cite{kane1998silicon} trapped ions,~\cite{haffner2008quantum} neutral atoms,~\cite{ludlow2015optical,henriet2020quantum} and dipolar molecules.~\cite{carr2009cold} The precise control of microwave wavelength, power, and polarization is of paramount importance to generate quantum superposition and entangled states in such systems with high fidelity. 

In particular the generation of microwaves with well-defined polarization becomes more and more important for quantum applications. Clean polarization ensures coupling between a well-defined pair of quantum states. While the generation of pure linear polarization is relatively straightforward, the creation of clean circularly polarized microwave fields is more challenging. The creation of purely circularly polarized microwaves has been critical for the recent demonstration of microwave shielding of ultracold dipolar molecules by our group and others,~\cite{karman2018microwave, karman2019microwave, anderegg2021observation, schindewolf2022evaporation, bigagli2023collisionally, lin2023microwave} which has motivated the design of the antenna presented in this work. Circularly polarized microwaves also play a critical role for the preparation of circular atomic Rydberg states,~\cite{raimond2001manipulating} which are expected to find applications as long-lived qubits,~\cite{facon2016sensitive,nguyen2018towards,cohen2021quantum} and for the control of hyperfine qubits in alkali atoms, for example for microwave clock operation,~\cite{bize2005cold} or to realize mid-circuit measurement protocols in advanced quantum algorithms.~\cite{graham2023mid} Beyond atomic and molecular systems, we expect that other quantum systems and qubits will also benefit from technologies to produce clean circularly polarized microwaves. 

Fundamentally, there are two approaches to create circularly polarized microwave fields: Designs based on a helical antenna naturally produce circular polarization as a result of the helix geometry.~\cite{kraus1949helical,Kraft1996MainbeamPP,kraus2002antennas} The handedness of the polarization is determined by the helicity of the antenna and cannot be tuned in situ. Also, helical antennas often require a reflector,~\cite{balanis2015antenna} which is not easy to tune and can hinder optical access. Alternatively, circular polarization arises by superimposing two linearly polarized microwave fields with an appropriate phase difference. This can be achieved using two orthogonal modes of a rectangular waveguide,~\cite{chen2023field} fields emerging from two microwave horns,~\cite{graham2023mid} or antenna arrays with discrete rotational symmetry.~\cite{signoles2014manipulations} Typical designs are often bulky and integration into quantum science setups, which often have limited space and other constraints, is challenging. 

In this letter, we describe the design and characterization of a microwave antenna array that consists of four loop antennas arranged in a cloverleaf shape. The array is planar, features high optical access, and the microwave polarization can be flexibly tuned by adjusting the relative phases between the constituent loops. Here, we optimize the array for left-circular polarization\footnote{The handedness is defined as left-handed (right-handed) if the electric field is rotating counterclockwise (clockwise) observed from the point of view of the source.} in the near field and characterize its properties by driving a rotational transition of sodium-cesium (NaCs) molecules, positioned 22 mm away from the antenna. We observe strong microwave coupling with a Rabi frequency of 2$\pi\times 46.1(2)$~MHz, corresponding to an electric field of 33(2) V/cm. In addition, we measure a high purity of circular polarization with an ellipticity of 2.3(4)$\degree$.

\begin{figure}
    \centering
    \includegraphics{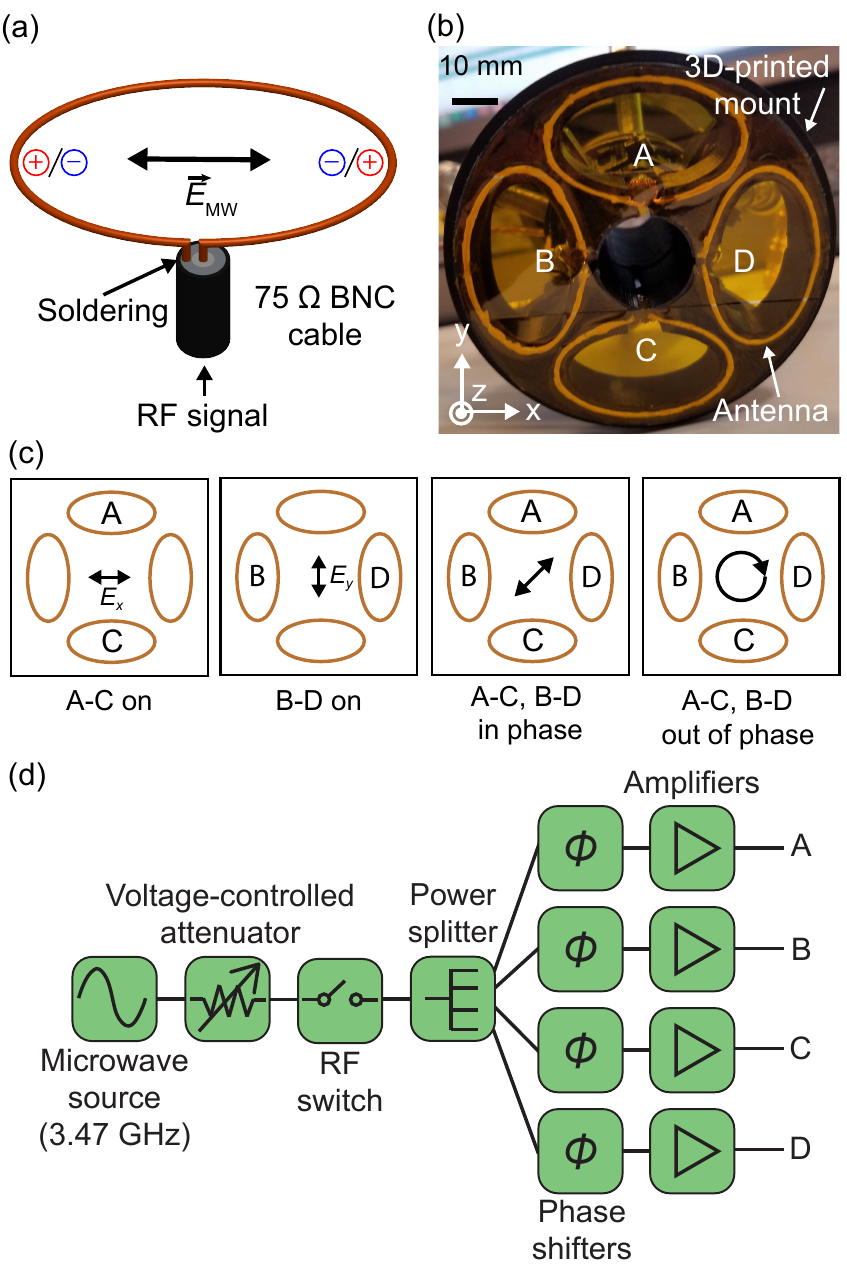}\\
    \caption{ Planar cloverleaf microwave antenna. (a) Drawing of a single loop. (b) Picture of the antenna array. A 3D-printed structure (black) serves as a mount. Each loop is wrapped in white Teflon tape for insulation. The bottom of the antenna mount is covered by Kapton tape to protect the viewport of the vacuum chamber that the antenna is mounted on. (c) Tunability of microwave polarization of the array. When pair A-C (B-D) is active, the electric field $E_x$ ($E_y$) is linearly polarized along the horizontal (vertical) axis. When A-C and B-D are operated in phase (out of phase), 45-degrees linear polarization (circular polarization) is generated. (d) Electronic setup supplying the cloverleaf antenna.
    }
    \label{fig:1}
\end{figure}

\section{Design}

Our antenna array is designed for an operating frequency of 3.47 GHz, corresponding to an in-vacuum wavelength of 86.5~mm. It consists of four elliptical loop antennas that are arranged in a cloverleaf shape, as shown in Fig.~\ref{fig:1}. Each loop has a semi-major axis of 16.7~mm and a semi-minor axis of 10~mm. The circumference is 86.5~mm. A 3D-printed mount made of PETG plastic holds the four loops in place (see Fig.~\ref{fig:1}(b)). A hole with a diameter of 20 mm in the middle of the mount allows for laser beams to propagate through the middle. The antenna array is less than 1~mm thick.

The individual loop antennas are made from 75~$\Omega$ coax BNC cables, following Ref.~\cite{miller2007studying} (see Fig.~\ref{fig:1}(a)). The jacket and the inner insulator are peeled off to expose the inner copper wire on a length that is slightly longer than one wavelength in vacuum. The copper wire is bent into an elliptical shape and its end is soldered to the braided metal shield of the coax cable to form a loop. The impedance of the resulting one-wavelength antenna in the array has a real part of 100~$\Omega$ with an additional imaginary contribution. To impedance match it to the 50~$\Omega$ microwave system, we employ a quarter-wavelength transformer and stub-tuning. The quarter-wavelength transformer changes the real part of the antenna impedance to $50$ $\Omega$; it is implemented by leaving the length of the unstripped part between the loop antenna and the BNC connector to be 5/4 of the wavelength in the 75 $\Omega$ BNC cable. An open circuit tuning stub cancels the imaginary part of the impedance; using a T-adaptor it is inserted between the 75 $\Omega$ cable and the 50 $\Omega$ cable that comes from the microwave amplifiers (see Fig.~\ref{fig:1}(d)). This setup suppresses reflections from the antenna by about 10 dB. We measure a bandwidth of 40 MHz for a single loop antenna.

Each loop generates a linearly polarized microwave field oscillating along its semi-major axis (see Fig.~\ref{fig:1}(a)). By combining the loops in different ways, the polarization of the array can be flexibly tuned (see Fig.~\ref{fig:1}(b) and (c)). The pair A-C (B-D) generates a microwave field that is linearly polarized along the $x$ ($y$)-axis, which we denote by $E_x$ ($E_y$) (see Fig.~\ref{fig:1}(c)). To maximize $E_x$ ($E_y$), we set the phase difference between loops A and C (B and D) to $\pi$. The polarization of the entire antenna array is controlled by the phase difference between $E_x$ and $E_y$. In order to create left (right)-circularly polarized microwave fields, we set the phase difference of $E_x$ relative to $E_y$ to be $-\pi/2$ ($+\pi/2$). In this work, we have optimized the antenna for left-circular polarization to maximize coupling to a $\sigma^{+}$-rotational transition of NaCs molecules (see Section IV).

The electronics stack supplying the cloverleaf antenna is shown in Fig.~\ref{fig:1}(d). A microwave generator (Rohde \& Schwarz SMA100B) produces a 3.47 GHz sine wave, whose amplitude can be tuned by a voltage-controlled attenuator (General Microwave D1954-OPT62) and switched by a RF switch (Mini-Circuits ZYSWA-2-50DR+). The signal is then split into four branches with a power splitter (Mini-Circuits ZN4PD1-63-S+). To tune the phases, we change the effective cable length between the power splitter and the amplifiers by inserting a series of commercial SMA adapters, which act as passive phase shifters. Then, each signal is amplified by a 15 W amplifier (Mini-Circuits ZHL-15W-422X-S+) and connected to one of the loop antennas.

\section{Numerical simulation}

\begin{figure*}
    \centering
    \includegraphics{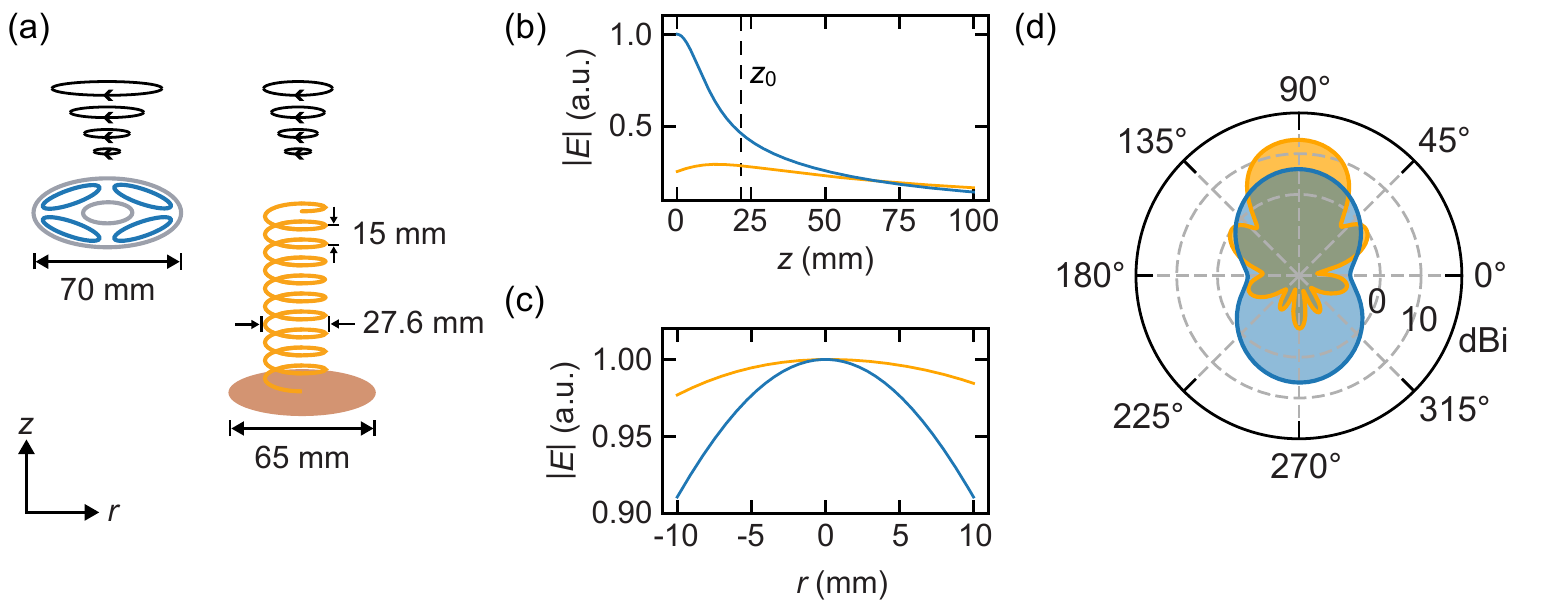}\\
    \caption{Simulation of the cloverleaf array and a reference helical antenna. Data in blue (orange) refers to the cloverleaf (helical) antenna. (a) Geometry of the antenna array and the helical antenna used in the simulation. (b) Amplitude of the total electric field along the axial direction. The vertical dashed line indicates $z_0 = 22$ mm, the distance where the antenna is experimentally used and characterized in Section IV. (c) Total electric field amplitude in the radial direction for $z_0$. The electric field amplitude is normalized to the value at position $r=0$ mm for both antennas. (d) Calculated directivity of both designs. The axial $z$-direction corresponds to $90^\circ$ in this plot.}
    \label{fig:2}
\end{figure*}

We numerically simulate the performance of the cloverleaf antenna, mostly focusing on near field operation. Using the Antenna Toolbox in Matlab,~\footnote{The Antenna Toolbox in Matlab employs the method of moments (MoM) and the results are crosschecked by running simulations on COMSOL, which uses the finite element method (FEM).} we calculate the radiation pattern generated by the antenna array. For comparison, we also consider a helical antenna.~\cite{balanis2015antenna} Helical antennas naturally produce circularly polarized microwaves when operating in axial mode. They have been widely used in atomic and molecular physics experiments,~\cite{ottl2006hybrid,boguslawski2019all,schindewolf2022evaporation,zheng2022compact,borowka2023continuous} making them a good standard for comparison. The parameters of the helical antenna are chosen to balance electric field amplitude and polarization purity. We assume a helix with 10 turns and a pitch spacing of 15 mm. The radius of the helix is 13.8 mm, the radius of the reflector disc is 32.4 mm, as shown in Fig.~\ref{fig:2}(a). 

As shown in Fig.~\ref{fig:2}(b)), for a given input power, the cloverleaf antenna has a stronger electric field amplitude than the helical antenna. At the specific distance of $z_0 = 22$ mm, relevant for the antenna characterization in Section IV, it is 1.6 times larger. This comes also with a larger electric field gradient for the cloverleaf antenna in the near field, while the two designs approach each other in the far field. Fig.~\ref{fig:2}(c) shows the radial amplitude profile for $z_0$. Compared with the helical antenna, the electric field amplitude of the cloverleaf array shows a stronger curvature in radial direction. 

Also, the directivity in the far field of the two designs shows a marked difference. In axial direction ($90\degree$) it is 6.2 dBi for the cloverleaf and 13.4 dBi for the helical antenna, indicating a higher directivity of the helical antenna. Given the symmetry of its design, the cloverleaf antenna emits symmetrically in the forward and backward direction. The electric field amplitude in the far field falls off quickly and near-field operation is favored. In addition, we calculate the reflection coefficient (S$_{11}$ parameters) for different microwave frequencies to quantify reflections from the antenna including the stub tuning.~\cite{ellingson2020electromagnetics} The calculation yields a voltage standing wave ratio (VSWR) 2:1 bandwidth of 80 MHz, which is compatible with the measured bandwidth of a single loop.

\section{Experimental characterization}

\begin{figure*}
    \centering
    \includegraphics{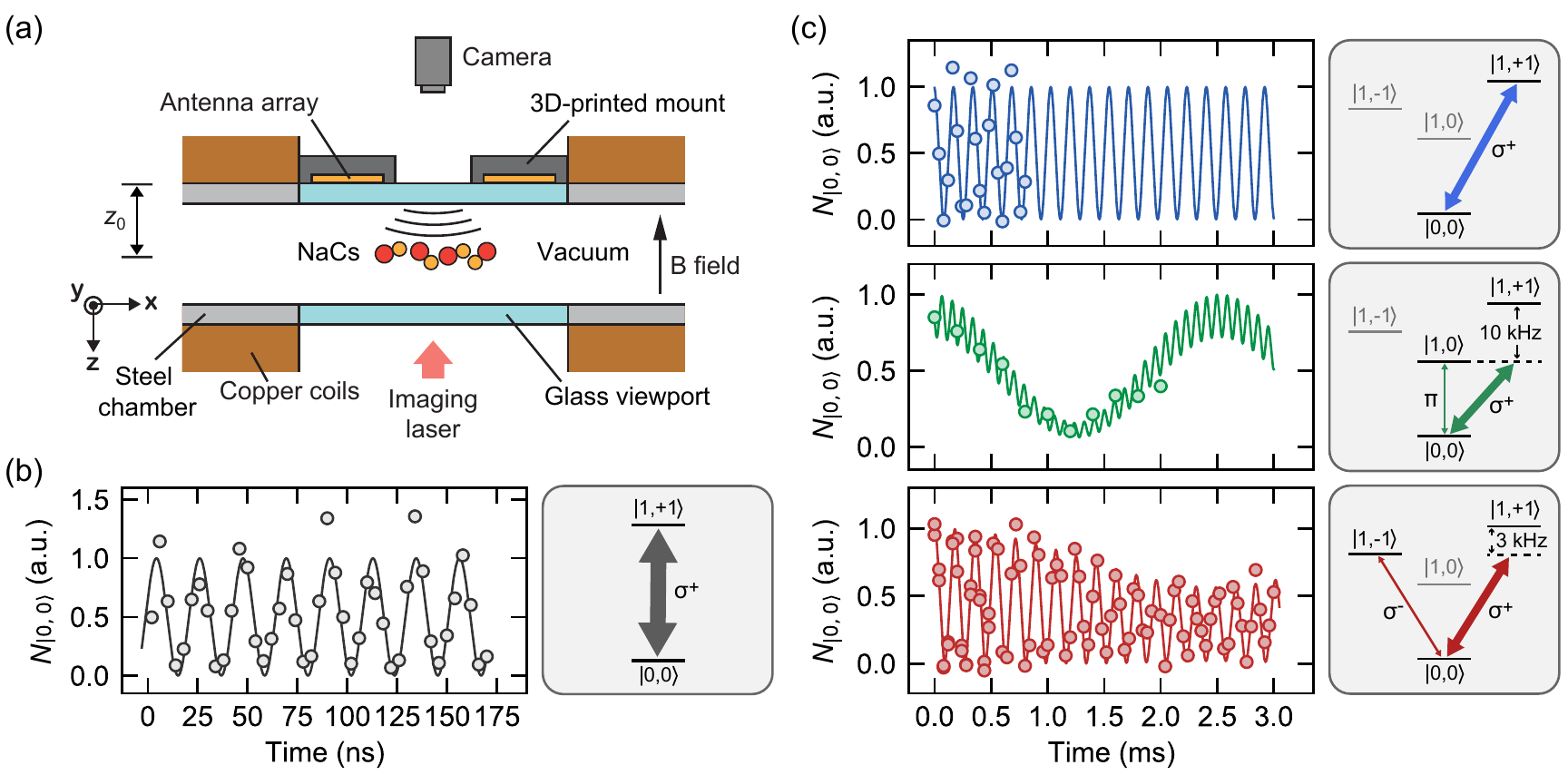}\\
    \caption{Characterization of the cloverleaf antenna with ultracold NaCs molecules. (a) The antenna array sits on top of a fused silica viewport of an ultrahigh vacuum chamber. The distance between the antenna and the molecules is $z_0=22$ mm or 0.26$\lambda$, where $\lambda$ is the wavelength of the resonant microwave field. (b) Fast Rabi oscillations on the $\sigma^{+}$-transition. The solid line is a sinusodal fit. (c) Slow Rabi oscillations on the $\sigma^{+}$ (upper panel), $\pi$ (middle panel), and $\sigma^{-}$-transition (lower panel) at low microwave power. The blue solid line is a sinusoidal fit to extract $\Omega_{+}$. To extract $\Omega_{\pi}$, a three-level model is employed to account for off-resonant coupling on the $\sigma^{+}$-transition manifested as fast jitter on top of the slow oscillation. The green solid line shows the three-level fit. For the $\sigma^{-}$-transition, a beat envelope is observed due to the interference between the resonant $\sigma^{-}$- and off-resonant $\sigma^{+}$-transition. The red solid line shows the three-level fit. }
    \label{fig:3}
\end{figure*}

We experimentally characterize the cloverleaf antenna in terms of electric field amplitude and purity of left-circular polarization. To this end we employ ultracold NaCs ground state molecules as an extremely sensitive quantum sensor. The antenna is mounted directly on the glass viewport of a stainless-steel vacuum chamber, $22$~mm away from the molecular sample, surrounded by copper solenoids, as illustrated in Fig.~\ref{fig:3}(a). Details on the preparation and quantum-state resolved detection of NaCs ground state molecules are reported in earlier work from our group.~\cite{warner2021overlapping, lam2022high, stevenson2023ultracold, warner2023efficient} In short, NaCs molecules are prepared in the vibrational and rotational ground state, $\ket{J, m_J } = \ket{0, 0 }$, where $J$ is the rotational quantum number and $m_J$ its projection on the quantization axis, which is defined by a magnetic field of 864 G along the vertical direction. From $\ket{0, 0}$, three excited rotational states, $\ket{1, -1}$, $\ket{1, 0}$, and $\ket{1, +1}$, can be accessed via an electric dipole transition at a frequency of 3.47 GHz. Left-circularly polarized microwave fields drive the $\sigma^+$- transition to $\ket{1, +1}$.

First, we measure the amplitude of the electric microwave field. To this end, we measure the Rabi frequency $\Omega$ of the microwave drive, which is related to the electric field amplitude via $E = \hbar \Omega/d_\mathrm{tr}$, where $d_{\rm tr}$ is the transition dipole moment. To record the data shown in Fig.~\ref{fig:3} (b), we snap on a resonant microwave field, let the molecules evolve under the microwave field for a variable time, snap the field off, and measure the population in state $\ket{ 0, 0}$. We observe unusally fast Rabi oscillations between states $\ket{ 0, 0}$ and $\ket{ 1, +1}$ with a Rabi frequency $\Omega / (2 \pi) = 46.1(2)$~MHz, indicating a strong microwave electric field. The transition dipole for the $\sigma^{+}$-transition is given by $d_\mathrm{tr} = d_{\rm NaCs}/ \sqrt{3}$, where $d_{\rm NaCs} = 4.75 (20)$~D is the permanent electric dipole moment of the NaCs molecules. ~\cite{dagdigian1972molecular} From this we extract an electric field amplitude of 33(2) V/cm at distance $z_0$.

Second, we characterize the purity of left-circular polarization. In the microwave frame, the polarization purity can be quantified by the ellipticity, defined by $\xi = \arctan (E^{\prime}_{-}/ E^{\prime}_{+})$, where $E^{\prime}_{-}/ E^{\prime}_{+}$ is the ratio of the electric field amplitude proportional to the ratio of Rabi frequencies of the $\sigma^{\pm}$-transitions in the microwave frame. In the lab frame, since the propagation direction of the microwave is not perfectly aligned with the quantization axis defined by the magnetic field, both $\sigma^{\pm}$- and a non-zero linear $\pi$-polarization components are measured. To this end, we tune the antenna into resonance with the transitions to $\ket{ 1, +1}$, $\ket{ 1, 0}$, and $\ket{ 1, -1}$ and measure the respective Rabi frequencies $\Omega_+$, $\Omega_{\pi}$, and $\Omega_-$. To achieve the high spectral resolution, we use low microwave intensity. 

We measure resonant Rabi oscillations between $\ket{0, 0}$ and $\ket{1, +1}$ with $\Omega_{+}$ = 2$\pi \times$ 5.8(4) kHz, as shown in the upper panel in Fig.~\ref{fig:3}(c). To obtain $\Omega_{\pi}$, we tune the microwave frequency on resonance with the $\ket{0,0}$ to $\ket{1, 0}$ transition, as shown in the middle panel of Fig.~\ref{fig:3}(c). Fitting the data with a three-level model, which takes into account off-resonant $\sigma^{+}$-coupling, we obtain $\Omega_{\pi}$ = 2$\pi \times$  0.38(7) kHz. Then, we tune the microwave frequency on resonance with the $\ket{0,0}$ to $\ket{1, -1}$ transition to extract $\Omega_{-}$, as shown in the lower panel of Fig.~\ref{fig:3}(c). Since the energy splitting between $\ket{1,+1}$ and $\ket{1, -1}$ is only 3 kHz, a beat envelope is observed in the data due to the simultaneous drive of $\sigma^{+}$ and $\sigma^{-}$ transitions. Fitting the data with a three-level model, we obtain $\Omega_{-}$ = 2$\pi \times$  0.23(2) kHz. From the ratio of Rabi frequencies, we calculate the ratio of electric field amplitudes $E_{-}/ E_{+}$ = 0.040(6) and $E_{\pi}/ E_{+}$ = 0.066(13) in the lab frame. Since the phase relations between the electric field components cannot be measured experimentally, we cannot determine the exact tilt angle between the propagation direction of the microwaves and the magnetic field, but we can estimate the tilt angle ranging from 5.1(7)$\degree$ to 5.5(7)$\degree$.~\footnote{For a given phase relation between the electric field components, we determine the Euler angles of the rotational transformation from the lab frame to the microwave frame by requiring the electric field component along the propagation direction of the microwave (or the linear $\pi$-polarization component) to be zero in the microwave frame. Then we rotate the electric field components to the microwave frame with this set of Euler angles and calculate the ratio $E^{\prime}_{-}/ E^{\prime}_{+}$ to extract the ellipticity. We scan the phase relations and obtain the range of tilt angles and the corresponding ellipticity. This procedure has been used and described in Ref.~\cite{schindewolf2022evaporation}} From this we infer the ratio of $E^{\prime}_{-}/ E^{\prime}_{+}$ in the microwave frame and obtain the ellipticity ranging from 2.1(2)$\degree$ to 2.4(2)$\degree$. Therefore, we conclude the ellipticity to be 2.3(4)$\degree$.

\section{Discussion}

Key features of the cloverleaf antenna are the compact form factor, the relatively high electric field amplitude in the near field, and the flexible tunability of the output polarization without the need to make physical changes to the antenna itself. This is especially useful for the correction of imperfections, such as reflections and field distortions from boundary conditions in the implementation environment. By tuning the relative phase between the loops, it is possible to switch the antenna from $\sigma^{+}$ to $\sigma^{-}$ polarization on demand. This degree of freedom is absent in helical antennas, where the polarization is set by the helicity of the spiral. We have demonstrated a small ellipticity in a challenging experimental environment with stainless steel and copper structures in the direct vicinity. In free space, likely even smaller ellipticity can be achieved. Here, we have implemented the antenna design for a resonance frequency of 3.47 GHz. It should be possible to adapt the design for microwave frequencies in a range from 1 to 20 GHz by adjusting the circumference of the individual loop antennas within practical limits. 

It should be possible to further improve the performance of the cloverleaf antenna with straightforward modifications. The electric field amplitudes produced by each elliptical loop likely differ by a small amount due to cross coupling between the loops, asymmetric reflections from the surrounding metal parts, imperfect manufacturing, or minor differences in the electronics stack of each loop. By adding amplitude control on the input of each elliptical loop, such inhomogeneities can be compensated and even finer control over the polarization purity could be achieved. Furthermore, it should be possible to increase the directivity of the cloverleaf antenna by adding a reflector that reflects backward radiation into forward direction. Our simulations show that this can increase the electric field amplitude by a factor of 1.6. Also, the inclusion of dynamical phase shifters in the electronics stack could further enhance the flexibility of the design. Finally, the form factor and thickness can be further reduced by implementing the antenna as a printed circuits board (PCB). This could also allow the direct integration into different experimental platforms, including setups in cryogenic environments, reaching beyond the use in atomic and molecular setups. 

\section{Conclusion}
We have presented a cloverleaf microwave antenna array with a high electric field amplitude and high polarization purity. It features a simple design, compact form factor, ease of manufacturing, and flexible fine tuning of polarization. In recent work, we have successfully used the antenna to realize collisionally stable NaCs ground state molecules, which enabled the first evaporative cooling of ultracold bosonic molecules.~\cite{bigagli2023collisionally} More generally, we expect our antenna design to facilitate control of molecular rotational states, which are expected to find uses in quantum simulation~\cite{bohn2017cold} and quantum computing~\cite{demille2002quantum, krems2009cold} thanks to long intrinsic coherence times.~\cite{gregory2023second} Also, the cloverleaf antenna may be useful for the implementation of quantum computing schemes with circular Rydberg atoms that require a low ellipticity $\xi < 2.5$$\degree$ to achieve high fidelity,~\cite{cohen2021quantum} which we have shown here to be within reach. While in this work the ellipticity was optimized for the electric field, as the system was used to drive an electric dipole transition, the polarization purity can also be optimized for magnetic field as relevant for magnetic dipole transitions. In addition, our design should be broadly adaptable for microwave quantum state control of atoms, from evaporative cooling of magnetically trapped atoms to the manipulation of atomic hyperfine qubits.~\cite{bluvstein2022quantum,graham2022multi} More generally, thanks to its flexibility, adaptability, and optical access, it should be useful for quantum systems that simultaneously require optical and microwave control, including, but not limited to, NV centers or trapped ions.

\section{Acknowledgement}
This work was supported by an NSF CAREER Award (Award No.~1848466), an ONR DURIP Award (Award No.~N00014-21-1-2721), and a Lenfest Junior Faculty Development Grant from Columbia University. C.W. acknowledges support from the Natural Sciences and Engineering Research Council of Canada (NSERC). W.Y.\ acknowledges support from the Croucher Foundation. I.S. was supported by the Ernest Kempton Adams Fund. S.W.\ acknowledges additional support from the Alfred P. Sloan Foundation.


%

\end{document}